\documentclass[letterpaper, 10 pt, conference]{ieeeconf}  

\IEEEoverridecommandlockouts                              

\usepackage{amssymb}
\usepackage{amsmath}
\usepackage{graphicx}
\usepackage{bm}
\usepackage{hyperref}
\usepackage{comment}

\usepackage{tikz}
\usepackage{tikz-3dplot}
\usetikzlibrary{calc}

\usepackage{algorithm}
\usepackage{algpseudocode}
\usepackage{soul} 
\usepackage{siunitx}

\usepackage[autostyle]{csquotes}
\MakeAutoQuote{‘}{’} 

\usepackage[backend=bibtex, doi=true,isbn=false,url=false,eprint=false, autolang=none, sorting=none]{biblatex}
\graphicspath{{figures/}}

\AtBeginBibliography{\small}

\definecolor{tum blue}{HTML}{0065BD}
\definecolor{tum blue 1}{HTML}{98C6EA}
\definecolor{tum blue 2}{HTML}{64A0C8}
\definecolor{tum blue 3}{HTML}{0073CF}
\definecolor{tum blue 4}{HTML}{005293}
\definecolor{tum blue 5}{HTML}{003359}

\definecolor{tum green}{HTML}{A2AD00}
\definecolor{tum orange}{HTML}{E37222}
\definecolor{tum ivory}{HTML}{DAD7CB}

\definecolor{tum dia violet}{HTML}{69085A}
\definecolor{tum dia dark blue}{HTML}{0F1B5F}
\definecolor{tum dia turquoise}{HTML}{00778A}
\definecolor{tum dia dark green}{HTML}{007C30}
\definecolor{tum dia light green}{HTML}{679A1D}
\definecolor{tum dia light yellow}{HTML}{FFDC00}
\definecolor{tum dia dark yellow}{HTML}{F9BA00}
\definecolor{tum dia dark orange}{HTML}{D64C13}
\definecolor{tum dia red}{HTML}{C4071B}
\definecolor{tum dia dark red}{HTML}{9C0D16}

\newcommand{\norm}[1]{\left\lVert#1\right\rVert}

\newcommand{\pdiff}[2]{\frac{\partial #1}{\partial #2}}

\newcommand\copyrighttext{%
	\footnotesize \textcopyright 2025 IEEE. Personal use of this material is permitted. Permission from IEEE must be obtained for all other uses, in any current or future media, including reprinting/republishing this material for advertising or promotional purposes, creating new collective works, for resale or redistribution to servers or lists, or reuse of any copyrighted component of this work in other works.
}  
\newcommand\copyrightnotice{%
	\begin{tikzpicture}[remember picture,overlay]  
		\node[anchor=south,yshift=10pt, xshift=0pt] at (current page.south) {\fbox{\parbox{\dimexpr\textwidth-\fboxsep-\fboxrule\relax}{\copyrighttext}}};  
	\end{tikzpicture}%
	\vspace{-0.35cm}  
}

\title{\LARGE \bf
Dynamic Constraint Tightening for Nonlinear MPC for Autonomous Racing via Contraction Analysis
}

\author{Joscha F. Bongard$^{1}$, Valentin L. Krieger$^{2}$, Boris Lohmann$^{1}$
\thanks{$^{1}$Chair of Automatic Control, Department of Engineering Physics and Computation, Technical University of Munich,
        Boltzmannstraße 15, 85748 Garching bei München, Germany
        {\tt\small \{joscha.bongard,lohmann\}@tum.de}}%
\thanks{$^{2}$
        {\tt\small krieger-valentin@web.de}}%
\thanks{*funded by the Deutsche Forschungsgemeinschaft (DFG, German Research Foundation) - 469341384}
}

\begin{document}

\maketitle
\thispagestyle{empty}
\pagestyle{empty}
\copyrightnotice{}

\begin{abstract}

This work develops a robust nonlinear Model Predictive Control (MPC) framework for path tracking in autonomous vehicles operating at the limits of handling utilizing a Control Contraction Metric (CCM) derived from a perturbed dynamic single track model. We first present a nonlinear MPC scheme for autonomous vehicles. Building on this nominal scheme, we assume limited uncertainty in tire parameters as well as bounded force disturbances in both lateral and longitudinal directions. By simplifying the perturbed model, we optimize a CCM for the uncertain model, which is validated through simulations at the dynamic limits of vehicle performance. This CCM is subsequently employed to parameterize a homothetic tube used for constraint tightening within the MPC formulation. The resulting robust nonlinear MPC is computationally more efficient than competing methods, as it introduces only a single additional state variable into the prediction model compared to the nominal scheme. Simulation results demonstrate that the homothetic tube expands most significantly in regions where the nominal scheme would otherwise violate constraints, illustrating its ability to capture all uncertain trajectories while avoiding unnecessary conservatism.

\end{abstract}


\section{Preliminaries}

\subsection{Introduction}

Autonomous driving is a dynamic field promising economic benefits and increased comfort while improving safety in critical situations as result of eliminating human error \cite{Betz2022}. Autonomous driving software is typically divided into the sub-tasks of environment perception and ego positioning, then object prediction and planning based on the predicted environment, and lastly control of the vehicle along a pre-defined path \cite{Betz2019SoftwareArchitecture}. The following focuses on the control part.

To make safe autonomous driving a real possibility, a sufficiently large set of possible scenarios need to be safely handled on part of the path controller. This includes problematic scenarios like emergency maneuvers at the handling limits of the vehicle.

While numerous competing controller schemes exist for autonomous vehicle control, Model Predictive Control (MPC) \cite{Rawlings2020MPC} emerged as the de-facto standard for high-performance schemes. MPC is uniquely suited for the task because of its ability to deliver near-optimal control for nonlinear systems under constraints. These advantages, under the assumption that the resulting optimal control problems (OCPs) can be solved sufficiently fast, promise to tackle the challenge of driving an autonomous vehicle safely and in near-optimal fashion even at the limits of handling.

One of the most important challenges in applying MPC lies in the highly problematic combination of state constraints and model uncertainty. This can lead to undesirable control behaviors when approaching constraints or even loss of recursive feasibility. To handle this problem, on the one hand stochastic MPC uses the propagation of probability distributions forward in time instead of a single trajectory and uses imposed chance constraints on this distribution. On the other hand, robust MPC only assumes upper bounds on the uncertainty and propagates convex sets forward in time, containing all uncertain trajectories. These sets are called \emph{tube}. The constraints are imposed on the tube spanning all possible trajectories. This is conceptually implemented by subtracting the tube size from the constraint bounds (\emph{Constraint Tightening} (CT)), thus allowing robust constraint satisfaction while still having the lower computational burden of optimizing over a single trajectory \cite{Rawlings2020MPC}.
A useful constraint tightening in the field of path tracking has no impact on behavior away from constraint bounds but enforces more conservative driving when uncertainties threaten to bring the vehicle to constraint bounds, and also reacts dynamically to the uncertainty in the present driving state to not compromise performance unnecessarily.

\begin{figure}[h!]
	\centering
	\includegraphics[scale=0.40]{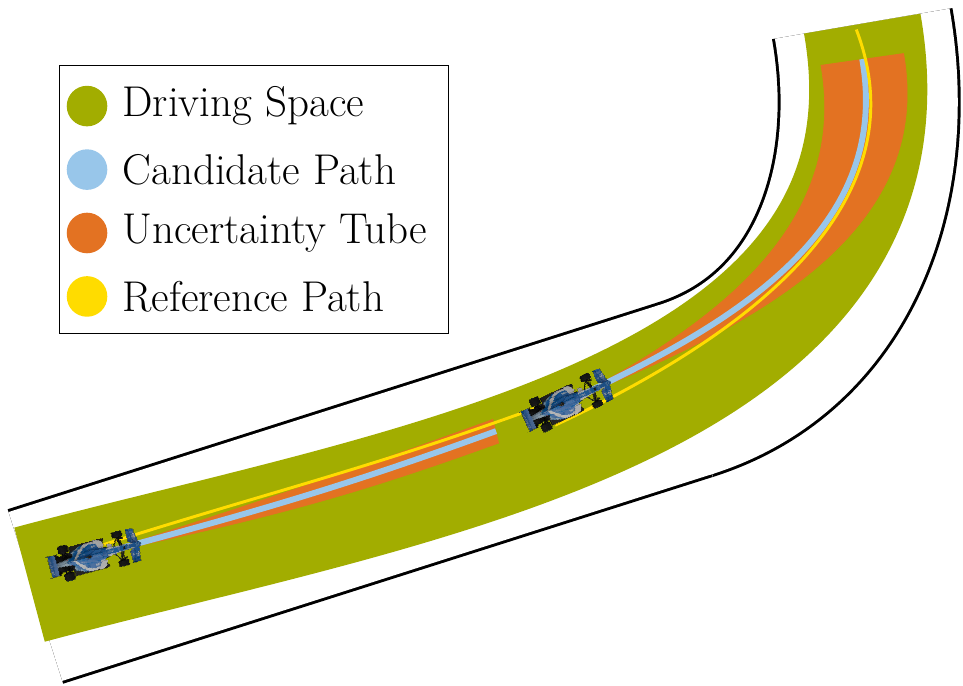}
	\caption{Visualization of homothetic uncertainty tube in different driving conditions}
	\label{fig:homothetic_tube}
\end{figure}

\subsubsection{Related Work}
Predictive path control of autonomous vehicles is a dynamic field. Recent overviews are given in \cite{Betz2022} and \cite{Stano2023}.

Tube MPC for linear models is a well-researched topic that yields good driving behavior even at high speeds given good low-level control is used. The robustification assists in avoiding states where the actual nonlinear driving behavior deviates too far from the linear design to handle \cite{Wischnewski2022TubeMPCApproachHighSpeedOvals}. For all its robustness, linear MPC in racing applications sacrifices performance and often struggles with vehicle stability in situations at the limits when nonlinearities become too relevant to ignore.

While nonlinear models increase the effort for parameter identification and computation, they provide the benefits of both increased safety at the limits and improved performance by better predicting actual vehicle behavior \cite{Stano2023}.
In autonomous racing, the tradeoff between model accuracy and low complexity usually demands single track models instead of simpler models, either outsourcing the tire model to low-level control \cite{Pierini2024} or incorporating it into the prediction \cite{Raji2022}.

Often nonlinear MPC schemes are combined with a constraint tightening based on the propagation of ellipsoids, where the actual distortion of the ellipsoids through nonlinear dynamics is approximated to obtain the robustness improvement. While these schemes are quite non-conservative, their high computational effort can conflict with the strict real-time requirements of autonomous racing \cite{Villanueva2017, Zarrouki2024}.

A nonlinear path-tracking MPC has previously been augmented with a rigid tube, where the tube stays constant over the prediction horizon, based on robust positively invariant sets \cite{Zieger2023RigidTubeMPC, Rawlings2020MPC}. While this approach is simpler than dynamic tubes, it degrades performance by being overly conservative especially at the early steps of the prediction horizon where it is advantageous to take advantage of the full acceleration potential of the vehicle.

Outside of autonomous racing, nonlinear MPC schemes have been successfully augmented with dynamic constraint tightening based on contraction metrics, especially for drones \cite{Sasfi2022, Singh16TubeBasedMPCAContractionTheoryApproach, Zhao2022a}, where the dynamics are simpler than for ground vehicles in that they don't involve the roughly non-holonomic behavior of tires.


\subsubsection{Contribution}
We use a nonlinear dynamic single track model to design a stabilizing nonlinear MPC scheme.
For this model with simplifications, assumed bounded parametric tire model uncertainty and force disturbances, we propose a novel method to compute a \emph{Control Contraction Metric}. The resulting Control Contraction Metric to the best of our knowledge is the first for this application and allows to quantify present uncertainty over the state space based on a physical model. This is used in the MPC to construct a non-conservative homothetic tube containing all uncertain trajectories under a manageable increase in computational burden. We show on a highly dynamic raceline that the MPC with tightened constraints is able to stabilize the vehicle at and even beyond the limits and prevents constraint violations and excessive performance degradation.

\subsubsection{Notation}
$\left[\right]_i$ denotes the $i$-th element of a vector, and $\left[\right]_{:,k}$ the $k$-th column of a matrix.
$\mathbb{I}_{\left[a,b\right]}$ denotes the set of integers between $a$ and $b$ while $\mathbb{R}_{\left[a, b\right]}$ denotes the closed interval of real numbers between $a$ and $b$.

\section{Methodology}

\subsection{Vehicle Model} \label{sec:vehicle_model}

The single track model (STM) presents a simple dynamic model for a car-like system in a flat plane. It disregards rolling motions and load transfer and lumps the front and rear axles each into a single tire. Details on the STM can be found in \cite{Jazar2008}, while this implementation builds mainly on \cite{Raji2022}.



\begin{figure}[h!]
	\centering
	\includegraphics[scale=0.94]{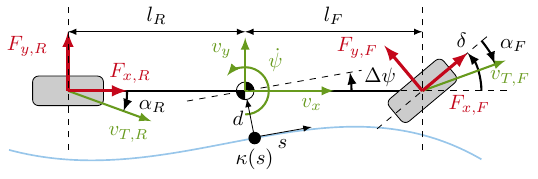}
	\caption{Dynamic single track model in curvilinear coordinates}
	\label{fig:STM}
\end{figure}

The model is expressed in curvilinear coordinates using the path length $s$, lateral deviation $d$ and relative heading $\Delta \psi$ to describe the vehicle position and orientation relative to the reference path. This path is characterized by its curvature $\kappa(s)$
A sketch of the dynamic STM in curvilinear coordinates is shown in Figure \ref{fig:STM}.

The state space model of the dynamic STM is given by

\begin{align} \label{eq:STM}
	\begin{bmatrix}
		\dot{s} \\
		\dot{d} \\
		\Delta\dot{\psi}\\
		\dot{v}_{x} \\
		\dot{v}_{y} \\
		\ddot{\psi}\\
		\dot{\delta}\\
		\dot{T}\\
		\dot{B}
	\end{bmatrix}
	=
	\begin{bmatrix}
		\left(v_{x} \cos \left(\Delta\psi\right)-v_{y} \sin \left(\Delta\psi\right)\right)\left(1-d \cdot \kappa\left(s\right)\right)^{-1} \\
		v_{x} \sin \left(\Delta\psi\right)+v_{y} \cos \left(\Delta\psi\right) \\
		\dot{\psi}-\kappa(s) \dot{s} \\
		a_x + v_{y} \dot{\psi} \\
		a_y - v_{x} \dot{\psi} \\
		\frac{1}{I_{z}}\left(F_{y,F} l_{F} \cos(\delta) + F_{x,F} l_{F} \sin(\delta) - F_{y,R} l_{R}\right)\\
		u_{d\delta}\\
		u_{dT}\\
		u_{dB}
	\end{bmatrix},
\end{align}
with accelerations at the center of gravity (CoG) in the vehicle body frame given by
\begin{subequations} \label{eq:accelerations}
\begin{align}
	a_x &= \frac{1}{m}\left( F_{x,R} + F_{x,F} \cos(\delta) - F_{y,F} \sin(\delta) - C_{r2} v_{x}^2 \right),\\
	a_y &= \frac{1}{m}\left( F_{y,R} + F_{y,F} \cos(\delta) + F_{x,F} \sin(\delta) \right).
\end{align}
\end{subequations}
For small steering angles, the slip angles at front ($\alpha_F$) and rear ($\alpha_R$) axle are approximated by 
\begin{subequations}
\begin{align}
	\alpha_{F} = \delta - &\arctan\left( \frac{v_y + l_{F} \dot{\psi}}{v_x} \right),\\
	\alpha_{R} = - &\arctan\left( \frac{v_y - l_{R} \dot{\psi}}{v_x} \right).
\end{align}
\end{subequations}
The lateral tire forces are given by the following force law based on a simplified \emph{Magic Formula} by Pacejka \cite{Pacejka2012TireAndVehicleDynamics}
\begin{align} \label{eq:pacejka_model}
\begin{split}
	F_{y,i} = F_{{z},i} \mathcal{D}_i &\sin (\mathcal{C}_i \arctan \left( \mathcal{B}_i \alpha_i \right)\\
	&- \mathcal{E}_i \left( \mathcal{B}_i \alpha_i - \arctan \left( \mathcal{B}_i \alpha_i \right) \right) ),
\end{split}
\end{align}
for $i \in \{{F, R}\}$, front and rear axles correspondingly, where the normal loads on the tires are assumed static and given by
$F_{{z},{F}} = l_{R}/(l_{F} + l_{R}) m g$ and
$F_{{z},{R}} = l_{F}/(l_{F} + l_{R}) m g$.

The front and rear longitudinal tire forces are given by a simplified model using throttle ($T$) and Brake ($B$) inputs
\begin{subequations}
\begin{align}
	F_{x, {F}} &= - C_{B,F} B - C_{r0}\\
	F_{x, {R}} &= C_{T} T - C_{B,R} B - C_{r0}.
\end{align}
\end{subequations}
with $T,B \in \mathbb{R}_{\left[0, 1\right]}$.
$m$ denotes the vehicle mass, $I_{z}$ the vehicle rotational inertia about the vertical axis, and $l_{F}, l_{R}$ are the front and rear distance from the respective wheel to the CoG. The gravity acceleration is denoted by $g$, and the coefficients $\mathcal{B}_{F/R}, \mathcal{C}_{F/R}$, $\mathcal{D}_{F/R}$, $\mathcal{E}_{F/R}$ denote the front and rear Pacejka coefficients.

Note that model \eqref{eq:STM} has been augmented with integrators at the inputs, i.e. the inputs are not $\delta, T, B$ directly but their derivatives. This is useful both for smoothing the inputs and for making the model input-affine, which simplifies computing a contraction metric later at the cost of increased relative degree.

The Pacejka model \eqref{eq:pacejka_model} captures the diminishing returns of the lateral forces the tires exert over their lateral slip angles as well as the negative returns once the force peak has been exceeded \cite{Pacejka2012TireAndVehicleDynamics}. This makes it suitable for controller design at the vehicle dynamic limits.

\subsection{Uncertainty Model} \label{sec:uncertainty_model}

\begin{figure}[h!]
	\centering
	\includegraphics[scale=1.0]{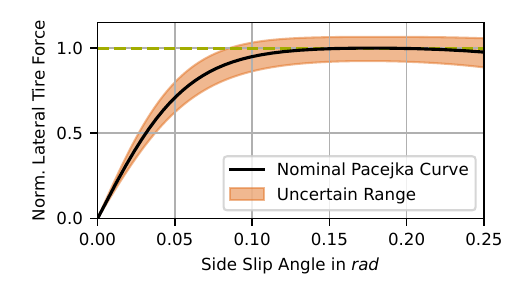}
	\caption{Sketch of tire curve uncertainty model}
	\label{fig:UncertModel1}
\end{figure}

The force-based nature of the dynamic STM \ref{eq:STM} renders the control strategy highly sensitive to inaccuracies in the estimation of lateral tire forces. To address this, the primary objective is to account for parametric uncertainties within the Pacejka tire model.
Specifically, additional uncertain lateral forces $\theta_1$ and $\theta_2$ at the front and rear axles are incorporated into the uncertainty framework, which are proportional to the respective lateral tire force. As visualized in Figure \ref{fig:UncertModel1} (for a higher uncertainty), this is equivalent to a variation of $\pm 2.5\%$ in the peak factor $\mathcal{D}$, that is proportional to the friction coefficient. The uncertainty interval is deliberately defined to be narrow in order to prevent excessively conservative behavior. 

Additionally, force disturbances acting on the vehicle's center of gravity are considered. While the disturbance $e_1$ is assumed to remain constant in the lateral direction, disturbance $e_2$ is dynamically linked to the brake and throttle commands in the longitudinal direction to account for external factors such as actuator delays, gear-shifting effects as well as incompatibility of simultaneous brake and throttle use. The disturbances are considered within a range of $\pm\SI{150}{\newton}$.

\subsection{Contraction} \label{sec:contraction}
\subsubsection{Contraction Theory}
This section builds on previous works \cite{Manchester2017, Zhao2022a, Sasfi2022} for restating contraction theory which is used in the following parametrization of the constraint tightening scheme.

Contraction analysis is a tool developed to quantify the velocity at which arbitrary trajectories of a system converge, inspired by fluid mechanics \cite{Manchester2017}. The contraction approach is comparable to classical Lyapunov theory but instead of dealing with the stability of definite equilibrium points, it deals with arbitrary reference trajectories which makes it suitable for autonomous driving, where the reference trajectory provided by a planning module can vary or even jump over time.

We assume an input-affine uncertain system of the form
\begin{align} \label{eq:f_w}
	f_w(x,u,\theta,e) = f(x) + B(x)u + G(x,u) \theta + E(x) e,
\end{align}
where $f(x)$ and $B(x)$ represent the nominal model, $\theta \in \Theta_0$ represents the constant but unknown parametric uncertainty in the uncertain parameter polytope $\Theta_0$ and $e \in \mathbb{E}$ the possibly time-varying unknown but bounded disturbance in the disturbance polytope $\mathbb{E}$ with according input matrices $G, E$ characterizing their interaction with the system.

While a \emph{Contraction Metric} gives a natural open-loop contraction of the system, it can be extended to cover non-open-loop contracting systems by considering a \emph{Control Contraction Metric} (CCM) \cite{Manchester2017}.
A CCM is a state-dependent matrix $M(x)$ giving an achievable contraction of the system under feedback, if
\begin{subequations}
\begin{align} \label{eq:ccm}
	\dot{M} + \left( A + B K \right)^\top M + M \left( A + B K \right) &\preceq - \beta M\\
	\underline{M} \preceq M(x) &\preceq \overline{M},
\end{align}
\end{subequations}
 with $\dot{M} = \sum_{j} \pdiff{M}{x_j}f_j$, where $f_j$ is the $j$-th element of $f$; $\underline{M},\overline{M} \succ 0$, with $A(x, u, \theta) = \pdiff{f}{x} + \pdiff{B}{x}u + \pdiff{G}{x}\theta + \pdiff{E}{x}e$ under input $u=k(x) + v$ with differential feedback $K=\pdiff{k}{x}$, where $v$ is a piecewise continuous signal.

We follow standard procedures to transform condition \eqref{eq:ccm} into a convex form by considering the \emph{dual} metric $W = M^{-1}$ \cite{Zhao2022a}

\begin{align} \label{eq:dual_ccm}
	-\dot{W} + A W + W A^\top + B Y + Y^\top B^\top \beta W \preceq 0,
\end{align}
with $\dot{W} = -M^{-1} \dot{M} M$, $Y = K W$.

\subsubsection{Control Contraction Metric for the STM}
This section presents a novel method to obtain a control contraction property from Section \ref{sec:contraction} of an uncertain STM.

To make solving \eqref{eq:dual_ccm} feasible, the model \eqref{eq:STM} is reduced to the dynamic states $v_x$, $v_y$, $\dot{\psi}$, $\delta$ and $C$, where $C$ represents a combined longitudinal force input acting on the center of gravity, replacing the longitudinal inputs $T$ and $B$. To handle the problem of the state- and input-dependent LMI \eqref{eq:dual_ccm}, we restrict the arguments of $W$ to only include $\begin{bmatrix}
	v_x, v_y, \dot{\psi}
\end{bmatrix}$ which characterize sets of stable driving states \cite{Erlien2013}. This model is then brought into the form \eqref{eq:f_w}.

According to the uncertain model \eqref{eq:f_w}, uncertainty input matrices $G$ and $E$ in \eqref{eq:f_w} define the influence of the parametric tire uncertainty $\theta = [\theta_1 , \theta_2]$ and external disturbance $e = [e_1, e_2]$ on the dynamics. Both $G$ and $E$ are chosen s.\,t. all elements of $\theta$ and $e$ are in $\mathbb{R}_{\left[-1, 1\right]}$ using the approximated dynamics of $v_x$, $v_y$:
\begin{align}
	G &= 
	\begin{bmatrix}
		-\frac{1}{m}\alpha_FF_{z,F}\sin{\delta}\cdot2.5\% & 0 \\
		-\frac{1}{m}\alpha_FF_{z,F}\cos{\delta}\cdot2.5\% & +\frac{1}{m}\alpha_RF_{z,R}\cdot2.5\% \\
		\frac{l_F}{I_z}\frac{1}{m}\alpha_FF_{z,F}\cos{\delta}\cdot2.5\% & -\frac{l_R \alpha_F}{m I_z}F_{z,R}\cos{\delta}\cdot2.5\% \\
		0 & 0 \\ 0 & 0
	\end{bmatrix}
	\\
	E &= 
	\begin{bmatrix}
		\frac{1}{m}C\cdot150 & 0 \\
		0 & \frac{1}{m}\cdot150 \\
		0 & 0 \\ 0 & 0 \\ 0 & 0\\
	\end{bmatrix}
\end{align}
For numerical reasons the side slip angles are over-approximated as a function of the yaw rate and lateral velocity.

For the polynomial approximation of \eqref{eq:f_w}, we constrain the maximum degree of polynomial basis functions to $4$. Note that it was not possible to find a constant metric, underscoring the need for a dynamic constraint tightening.

The subset of the state space in which \eqref{eq:dual_ccm} must hold is restricted by box constraints on every state and input as well as a safe driving envelope characterizing the relation between the side slip angle and yaw rate, inspired by \cite{Erlien2013}. 

To satisfy polynomial requirements of sum of squares (SoS) Programming \cite{Parillo2000}, the tire curves \eqref{eq:pacejka_model} are approximated linearly, the trigonometric functions in the model are approximated using Chebyshev polynomial approximations up to order $3$, and both the inverse tangent and scalar inverse functions are approximated linearly.

To obtain a finite-dimensional optimization problem, we use a formulation of the contraction conditions \eqref{eq:dual_ccm} as Sum of Squares Programming \cite{Parillo2000}. This formulation has the advantage that at least for the polynomial approximation, the solution yields global guarantees within the given subset of states and inputs. For the computation of a dual metric \eqref{eq:dual_ccm}, a MATLAB code inspired by \cite{Sasfi2022} is developed, that solves the given problem in \texttt{YALMIP} \cite{Löfberg2004} using the \texttt{MOSEK} \cite{mosek} solver.

Due to the complicated state space representation, the overall initial contraction rate has to be selected small to obtain a CCM. However in reality a part of the state combinations is not realistic. Therefore contraction conditions are checked using numerous optimistic trajectories and extreme scenarios taken from simulation, where the related contraction rate is calculated using eigenvalue analysis \cite{Sasfi2022}.
In the following all relevant constants and expressions for the construction of the tube size dynamic \eqref{eq:f_sigma} are derived. As the orthogonal deviation $d$ is not directly incorporated, relevant constants are reconstructed from the included variables for worst-case conditions.

\subsection{MPC Setup} \label{sec:mpc_setup}

The complete nominal OCP for the MPC scheme is given by \eqref{eq:mpc_ocp}.
It builds on previous works, with the model from Section \ref{sec:vehicle_model}, while costs and constraints are inspired by \cite{Wischnewski2022TubeMPCApproachHighSpeedOvals}.

\begin{subequations} \label{eq:mpc_ocp}
\begin{align}
	\min \sum_{k=0}^{N-1} y_k^\top Q y_k + u_k^\top R u_k + y_N^\top Q_N y_N \label{eq:mpc_cost_overall}\\
	y_k = \begin{bmatrix}
		d_k \\
		\dot{d}_k \\
		v_k - v_{ref,k}
	\end{bmatrix}, \,
	u_k = \begin{bmatrix}
		u_{d\delta}\\
		u_{dT}\\
		u_{dB}
	\end{bmatrix}, \,
	y_N = \dot{d}_N \label{eq:mpc_cost}\\
	x_{k+1} = f_d(x_k, u_k), \label{eq:mpc_dynamics}\\
	d_{\mathrm{min}} \leq d \leq d_{\mathrm{max}} \label{eq:constr_d}\\
	\delta_\mathrm{min} \leq \delta \leq \delta_\mathrm{max} \label{eq:constr_delta}\\
	0 \leq T \leq 1, \quad 0 \leq B \leq 1 \label{eq:constr_TandB}\\
	u_{\delta,\min} \leq u_\delta \leq u_{\delta,\max} \label{eq:constr_u_delta}\\
	u_{T,\min} \leq u_T \leq u_{T_,\max} \label{eq:constr_u_T}\\
	u_{B,\min} \leq u_{B} \leq u_{B,\max} \label{eq:constr_u_B}\\
	a_x + m_r a_y \leq C_r,\quad r \in \mathbb{I}_{\left[1, 8\right]} \label{eq:constr_gg_diagram}\\
	v_{N} \leq v_{\mathrm{planner}, N} \label{eq:constr_v_terminal}
\end{align}
\end{subequations}

The discrete-time prediction model $f_d(x_k, u_k)$ is obtained by discretizing the continuous-time nominal dynamics \eqref{eq:STM}.

In \eqref{eq:mpc_cost_overall}, \eqref{eq:mpc_cost}, \eqref{eq:mpc_dynamics}, the index $k$ indicates the discretized version of the corresponding continuous-time signal.

To imitate a longer prediction horizon, the terminal cost in \eqref{eq:mpc_cost} incentivizes aligning the terminal speed vector to the reference.

Constraints \eqref{eq:constr_d} to \eqref{eq:constr_gg_diagram} are imposed throughout the prediction horizon.

The terminal constraint \eqref{eq:constr_v_terminal} ensures that the controller keeps the car under the safe terminal speed requested by a planner module to ensure a potential emergency trajectory at the dynamic limits following the current trajectory can be safely tracked.

Since the Pacejka model \eqref{eq:pacejka_model} only captures nonlinearity in the lateral tire force, we add the combined friction limit constraint \eqref{eq:constr_gg_diagram}. This constrains the combined body frame accelerations \eqref{eq:accelerations} at the CoG to lie within a convex octagon to prevent excessive combined accelerations.

In order to keep OCP \eqref{eq:mpc_ocp} feasible under all possible circumstances, constraints \eqref{eq:constr_d} and \eqref{eq:constr_gg_diagram} are relaxed using slack variables with linear and quadratic norm costs.

\subsection{Robustification}
This section builds upon previous theoretical work \cite{Sasfi2022} and gives implementation details.
The contraction analysis from Section \ref{sec:contraction} under the uncertainty model from Section \ref{sec:uncertainty_model} is used to parameterize a simple constraint tightening scheme for OCP \eqref{eq:mpc_ocp} as proposed in \cite{Sasfi2022}. The constraints \eqref{eq:constr_d} to \eqref{eq:constr_gg_diagram} are formulated as one-sided constraints of the form
\begin{align}
	h_j(x,u) \leq 0, \quad j \in \mathbb{I}_{[1, r]},
\end{align}
that can be easily tightened. The tightened constraints read as
\begin{align} \label{eq:constraint_tightening}
	h_j(x,u) + c_j \sigma \leq 0,
\end{align}
where $c_j > 0$ are tightening constants and $\sigma$ is the time-dependent \emph{tube size} with the dynamics
\begin{align} \label{eq:tube_dynamics}
	\dot{\sigma} = - \left(\beta - L_\mathbb{E} \right) \sigma + f_\sigma(x, u),
\end{align}
that is directly incorporated into the MPC prediction model. When properly parametrized, the CT \eqref{eq:constraint_tightening} covers all uncertain trajectories in a tube \cite{Sasfi2022}. Here, $\beta$ refers to the continuous-time contraction rate from \eqref{eq:ccm}, $L_\mathbb{E}$ is the disturbance Lipschitz bound and $f_\sigma$ the assumed uncertainty with
\begin{align}
\begin{split}
	f_\sigma = \max_{\substack{i \in \mathbb{I}_{\left[1, n_{\theta_0}\right]}\\
			j \in \mathbb{I}_{\left[1, n_{\mathbb{E}}\right]}}}\left(\sum_{k=1}^p L_{\mathrm{G}, k}\left|\left[\theta^i-\bar{\theta}\right]_k\right| \sigma + f_{\sigma,\mathrm{dyn}} \right), \label{eq:f_sigma}
\end{split}
\end{align}
with
\begin{align} \label{eq:f_sigma_dyn}
	f_{\sigma,\mathrm{dyn}} = \norm{G\left(x, u\right)\left(\theta^i-\bar{\theta}\right)+E\left(x\right) e^j}_{M(x)}.
\end{align}
$n_{\theta_0}$ numbers the vertices $\theta^i$ of $\Theta_0 \subset \mathbb{R}^p$, while $n_{\mathbb{E}}$ numbers the vertices $e^j$ of $\mathbb{E}$.
All relevant constants and expressions $c_j$, $L_\mathbb{E}$ and $L_{G,k}$ can be directly derived from the CCM offline. We refer to \cite{Sasfi2022} for their calculation.

%

%
%
\subsection{Controller Implementation}
The controller is implemented based on the MPC software framework \texttt{acados} \cite{Verschueren2021acados}. The continuous-time problem is discretized using a $4$th order Runge-Kutta method with a step size of \SI{67}{\milli\second} at a horizon length of $36$ steps. This gives a practically proven horizon length of around \SI{2.4}{\second} \cite{Stano2023} while keeping solving times quick.


To make OCP \eqref{eq:mpc_ocp} computationally feasible, we use the \emph{real-time iteration} \cite{Diehl2002RTI}, where the nonlinear program is solved approximately by doing a single Sequential Quadratic Programming iteration around the previous solution. To solve the resulting Quadratic Program, we use the high-performance interior-point solver \texttt{HPIPM} \cite{Frison2020} with a Gauss-Newton hessian approximation.

\subsection{Simulation Results}
\subsubsection{Simulation Algorithm}
To highlight the benefits of enhancing the MPC scheme with the proposed dynamic constraint tightening, an idealized racetrack simulation is conducted. Emulating the real-world application of the control algorithm, an offline-generated global raceline of the \emph{Yas Marina Circuit} is used as reference.
The simulation algorithm is as follows:
\begin{enumerate}
	\item Determine the vehicle's position relative to the global raceline.
	\item Extract the relevant segment of the global raceline for the finite prediction horizon $N$.
	\item Solve the OCP \eqref{eq:mpc_ocp}
	\item Apply the first step of the computed optimal input sequence, return to step 1.
\end{enumerate}
A demanding global raceline is specifically designed to challenge the vehicle at its handling limits.
To evaluate the impact of constraint tightening, a variation in tire parameters is introduced in the simulation model only, by modifying the Pacejka coefficients: reducing $\mathcal{B}$, $\mathcal{C}$, and $\mathcal{D}$ by $5\%$, while simultaneously increasing $\mathcal{E}$ by an equivalent percentage. This shift in the Pacejka tire curve simulates a realistic scenario of reduced lateral grip. Additionally, a constant force disturbance of \SI{200}{\newton} is applied to the vehicle's center of gravity in both the longitudinal and lateral directions.
\subsubsection{Dynamic Tube Size Parameterization}

\begin{figure}[h!]
	\centering
	\includegraphics[scale=1.0]{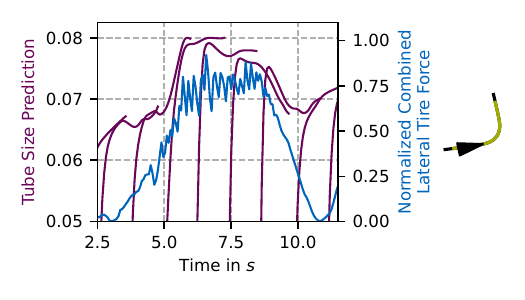}
	\caption{Predicted tube sizes and lateral tire grip demand at selected sampling points over the finite horizons during the first corner at \emph{Yas Marina Circuit} (Sketch on the right)}
	\label{fig:STMRes1}
\end{figure}

Figure \ref{fig:STMRes1} illustrates the tube size predictions according to \eqref{eq:tube_dynamics} across the respective horizon for selected sampling points while navigating a typical hairpin turn. The tube size parametrization is influenced by lateral tire grip demand, throttle/brake commands, and velocity, with its primary determinants being the CCM and the uncertainty and disturbance model \eqref{eq:f_sigma}. Among these factors, the lateral tire grip demand stands out as one of the largest influences on the tube dynamics, as tire uncertainty critically affects control accuracy. By progressively tightening constraints based on the tube size, the system achieves enhanced adaptability through increased flexibility in the initial states, thereby improving overall performance compared to rigid tube schemes.

\subsubsection{Slip Test}
To simulate unexpected and challenging conditions, the grip potential is reduced at the rear axis along a part of the path by saturating the force in lateral and longitudinal direction at $50\%$ of the nominal maximal value in the same scenario as shown in Figure \ref{fig:STMRes1}.

\begin{figure}[h!]
	\centering
	\includegraphics[scale=1.0]{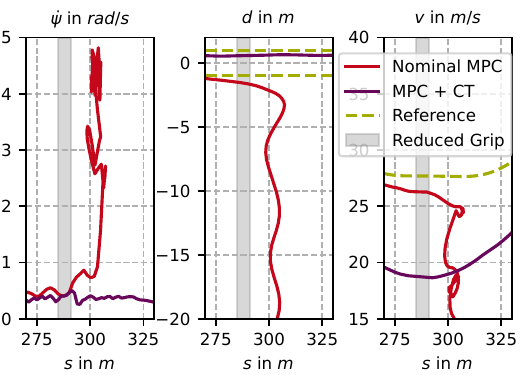}
	\caption{Yaw rate $\dot{\psi}$, orthogonal path deviation $d$ (Limits green dashed) and total velocity $v$ for the progress along the path $s$) during slip test in first corner of \emph{Yas Marina Circuit} as in Figure \ref{fig:STMRes1} using the nominal and robustified algorithm}
	\label{fig:STMRes4}
\end{figure}

Figure \eqref{fig:STMRes4} shows the results of the slip test for the MPC both with and without the proposed constraint tightening (CT). The area of reduced grip is indicated by the gray area. The yaw rate momentarily increases in both cases but since the robustified controller drives at a lower speed with improved uncertainty rejection potential, it is able to quickly recover, while the nominal controller tracks the speed more closely and therefore is driving at already saturated tires, entering a spin originally caused by the reduced grip from which it does not recover. 
This indicates that the robustified controller can cope with greater uncertainty than it was designed for in Section \ref{sec:uncertainty_model}.

\subsubsection{Influence of Constraint Tightening on Constraint Violations}

\begin{figure}[h!]
	\centering
	\includegraphics[scale=1.0]{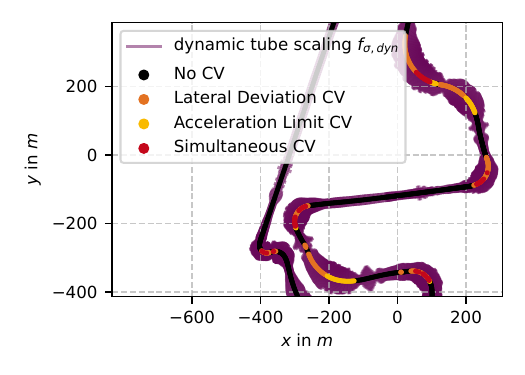}
	\caption{Visualization constraint violations in nominal MPC and assumed uncertainty \eqref{eq:f_sigma_dyn} in simulation at \emph{Yas Marina Circuit}}
	\label{fig:STMRes2}
\end{figure}

Constraint tightening ensures a remaining safety margin in lateral tire grip potential, which is dynamically scaled according to the robust stabilizability of the current driving maneuver by reducing speed, automatically making tradeoffs to ensure vehicle stability. As a result, tire parameter uncertainty is successfully rejected and no constraint violations (CV) occur, as the control algorithm adapts to model mismatch by adjusting the operating point within the Pacejka curve. Figure \ref{fig:STMRes2} shows that the dynamic part of the tube \eqref{eq:f_sigma_dyn} tends to be larger where the nominal scheme violates constraints while is small and therefore non-conservative where constraint satisfaction is not endangered.

Notably, the algorithm demonstrates the ability to handle significantly larger variations in Pacejka coefficients than those for which it was originally designed, highlighting the inherent conservatism in the CCM design. In contrast, grip overestimation in case of the nominal controller leads to severe infeasibility issues and constraint violations in both orthogonal deviation and acceleration limits.

\subsubsection{Path Tracking Performance}

\begin{figure}[h!]
	\centering
	\includegraphics[scale=1.0]{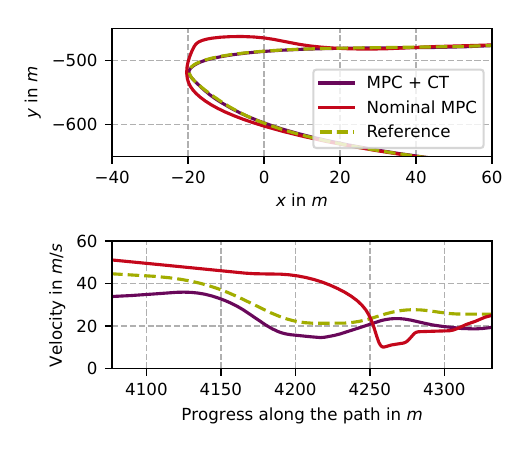}
	\caption{Performance of nominal and robustified MPC under uncertainty}
	\label{fig:STMRes3}
\end{figure}

Figure \ref{fig:STMRes3} illustrates the trajectory and velocity profiles for both the nominal and robustified control cases during navigation of a challenging corner at \emph{Yas Marina Circuit}. The robustified approach demonstrates more conservative behavior in scenarios where stability may be jeopardized, enabling rapid correction of model mismatch caused by uncertainties. Consequently, this approach enhances vehicle stability and therefore also path-tracking accuracy by automatically making compromises at the expense of velocity tracking. In contrast, the nominal approach struggles under uncertainty, particularly in sharp corners where tire capabilities are overestimated. This results in the optimization problem encountering feasibility issues at the limits, leading to significant degradation in optimization performance, large deviations from the intended path, and inappropriate velocity profiles. While the robustified algorithm effectively avoids unstable driving states (e.g. drifting) and demonstrates the ability to recover from such states if they occur, uncertainty frequently induces instability in the nominal case, increasing the risk of loss of vehicle control.

\subsubsection{Computational Effort}

\begin{table}[h!]
	\centering
	\renewcommand{\arraystretch}{1.2} 
	\begin{tabular}{l|p{1.5cm}|p{1.5cm}}
		MPC method & Med Solving & $P_{99}$ Solving \\
		\hline
		Nominal & \SI{1.861}{\milli\second} & \SI{3.810}{\milli\second} \\
		w/ Constraint Tightening & \SI{2.248}{\milli\second} & \SI{3.459}{\milli\second} \\
	\end{tabular}
	\caption{Median and $99$th Percentile Solving Time Comparison of MPC Implementations}
	\label{tab:mpc_timings}
\end{table}

Table \ref{tab:mpc_timings} shows the computation times for solving \eqref{eq:mpc_ocp} both with and without the presented constraint tightening during the lap of the \emph{Yas Marina Circuit} from Figure \ref{fig:STMRes2}. Table \ref{tab:mpc_timings} shows that the median solving time increases approximately $25\%$ compared to the nominal implementation, but remains within an acceptable range. This moderate increase can be attributed to most of the computational effort, i.e. the CCM condition \eqref{eq:dual_ccm} being made offline. The online effort is only moderately increased by the scalar tube dynamics \eqref{eq:tube_dynamics}. Notably, in the presence of uncertainties and disturbances, the 99th percentile ($P_{99}$) of solving times is even lower for the robust scheme, as the nominal algorithm frequently encounters numerical issues at the constraint bounds that complicate the optimization. Based on these tests and the given $P_{99}$ as the worst case, a control frequency of \SI{288}{\hertz} could be achieved on similar hardware in real-time iterations using the proposed dynamic CT.

All simulations were performed on a desktop computer equipped with a 2.8GHz Intel i7-1165.G7 processor.

\section{Discussion}

The presented robustification method is flexible in the sense that it can be applied to MPC schemes of various costs and prediction models, including path tracking as well as contouring control schemes. While the presented constraint tightening scheme is more conservative than some competing methods, it only requires a moderate increase in computation by offloading a major part of the effort offline, while still being able to dynamically react to present uncertainty online.

In general, nonlinear MPC has the remarkable property of being able to automatically make trade-offs and recover from overstepping handling limits such as spins. While this benefit can be jeopardized under uncertainty, the presented constraint tightening can help to preserve it without resorting to overly conservative behavior.

One downside of nonlinear MPC schemes is the increased requirement of accurate parametrization compared to simpler schemes. While still present, the proposed constraint tightening helps mitigate the susceptibility to parameter errors.

Compared to schemes with rigid tubes, the presented scheme offers better performance by keeping the complete potential at the first step of the optimization problem and only progressively tightens subsequent steps based on the predicted driving state.

An implementation of the presented algorithm on a full-scale racecar is currently being worked on as part of the team \emph{TUM Autonomous Motorsports} \cite{Betz2022TUMAutonomousMotorsport}.





%

\printbibliography

\end{document}